\documentclass[12pt, ams, onecolumn, amssymb, floatfix, longbibliography]{revtex4-2}

\usepackage{amsmath}
\usepackage{color}
\usepackage{amsfonts}
\usepackage{amssymb}
\usepackage{graphicx}
\usepackage{geometry}
\usepackage{hyperref}
\usepackage{comment}
\usepackage{tabularx}
\usepackage{bm}
\usepackage{euscript}
\usepackage{graphicx}
\usepackage{color}
\usepackage{amsfonts}
\usepackage{exscale}
\usepackage{amsbsy}
\usepackage{textcomp}
\usepackage{comment}
\usepackage{hyperref}
\usepackage{slashed}
\usepackage{mathtools}
\usepackage{tabularx}
\usepackage{bm}
\usepackage{euscript}
\usepackage{graphicx}
\usepackage{color}
\usepackage{amsfonts}
\usepackage{exscale}
\usepackage{amsbsy}
\usepackage{subfig}
\usepackage{textcomp}
\usepackage{comment}
\usepackage{hyperref}
\usepackage[dvipsnames]{xcolor}
\usepackage{natbib}
\usepackage{enumitem}

\usepackage{ulem}

\numberwithin{equation}{section}

\begin{document}

\title{Weiss Oscillations in the Galilean-Invariant Dirac Composite Fermion Theory for Even-Denominator Filling Fractions of the Lowest Landau Level}
\bigskip
\author{Yen-Wen Lu$^1$}
\author{Prashant Kumar$^2$}
\author{Michael Mulligan$^1$}
\affiliation{$^1$Department of Physics and Astronomy, University of California, Riverside, California 92511, USA}
\affiliation{$^2$Department of Physics, Princeton University, Princeton, New Jersey 08544, USA}

 \bigskip
 \bigskip
 \bigskip
 \bigskip
 \bigskip
 \bigskip

\begin{abstract}
Standard field theoretic formulations of composite fermion theories for the anomalous metals that form at or near even-denominator filling fractions of the lowest Landau level do not possess Galilean invariance. To restore Galilean symmetry, these theories must be supplemented by ``correction" terms. We study the effect of the leading ``correction" term, known as the dipole term, in the Dirac composite fermion theory (a theory that consists of a Dirac fermion coupled to an Abelian Chern-Simons gauge field) on quantum oscillations in the electrical resistivity due to a periodic scalar potential about even-denominator filling fractions. We find the dipole term to be insufficient to resolve the systematic discrepancy, discovered in Kamburov $et$ $al$. [Phys.~Rev.~Lett.~{\bf 113}, 196801 (2014)], between the locations of the oscillation minima predicted by Dirac composite fermion theory without Galilean invariance and those observed in experiment. Further, in contrast to Hossain $et$ $al$., [Phys.~Rev.~B {\bf 100}, 041112 (2019)], we find the quantum oscillations about the half-filled and quarter-filled lowest Landau level to have qualitatively similar behavior. This analysis uses a mean-field approximation, in which gauge field fluctuations are neglected. Based on this and previous analyses, we speculate that the discrepancy with experiment may be an indirect signature of the effect of gauge field fluctuations in composite fermion theory.
\end{abstract}

\maketitle

\bigskip

\newpage

\thispagestyle{empty}

\tableofcontents

\newpage

\setcounter{page}{1}

\section{Introduction}

\label{introduction}

Normal metals exhibit quantum oscillations \cite{abrikosovmetals1988}; so too do certain anomalous metals (e.g., Ref. \cite{doi:10.1098/rsta.2010.0243}).
Here we consider the anomalous metals that form when a 2d system of interacting electrons is at or near even-denominator filling fractions $\nu=1/2m$ (for $m$ a positive integer) of the lowest Landau level \cite{PhysRevB.40.12013, PhysRevB.46.10468}.
Quantum oscillations about $\nu=1/2m$ suggest a Fermi surface of low-energy excitations that couple to an effective magnetic field $B^\ast \equiv B - 2 m \Phi_0 n_e$ ($B$ is the external magnetic field, $\Phi_0 = hc/e$, and $n_e$ is the electron density) \cite{Willett97, shayegan2020probing}.
In particular, oscillations in the electrical resistivity (known as Weiss oscillations \cite{Weissfirst, gerhardtsweissklitzing}) occur when a unidirectional periodic scalar potential is applied and $B$ is varied about $\nu=1/2m$ (say, at fixed $n_e$).
About $\nu=1/2$, the oscillation minima are found at magnetic fields $B_p$ satisfying
\begin{align}
\label{weissformula}
\ell^2_{B^\ast_p} = {a \over 2 k^\ast_F} \big(|p| + {1 \over 4} \big), \quad p = \pm 1, \pm 2, \pm 3, \ldots.
\end{align}
Here, $\ell_{B^\ast_p} = \sqrt{\hbar c/e|B^\ast_p|}$ is the magnetic length for effective magnetic field $B^\ast_p = B_p - 2 \Phi_0 n_e$ at the $p$-th oscillation minima, $a$ is the period of the applied potential, and the sign of $p$ equals the sign of $B_p^\ast$.
(For the range of $n_e$ and $a$ considered experimentally, the most robust oscillation minima occur closer to $\nu=1/2$ than conventional Shubnikov-de Haas oscillations, which are generally obscured by fractional quantum Hall states.)
Prior to 2014, only the first few minima $|p| \leq 2$ were resolvable and a good fit to the data could be achieved by taking $k_F^\ast = \sqrt{4 \pi n_e}$ \cite{Willett97, shayegan2020probing}. 
This and similar results \cite{jainCF, Willett97} help to form the phenomenological justification for the composite fermion theory of the $\nu = 1/2$ state due to Halperin, Lee, and Read (HLR) \cite{halperinleeread}.
In this theory, the $\nu=1/2$ state is described by a collection of nonrelativistic ``composite" fermions interacting via a Chern-Simons gauge field, with $k_F^\ast$ identified as the composite fermion wave vector.

It came, then, as a surprise when, due to improvements in sample quality and experimental design, Kamburov $et$ $al$.~\cite{Kamburov2014} reported oscillation minima that were no longer symmetrically distributed about $\nu=1/2$ (at fixed $n_e$ and varying $B$).
The data can only be fit using Eq.~\eqref{weissformula} if $k_F^\ast$ now varies with $\nu$:
\begin{align}
\label{kamburovresult}
k_F^\ast =
\begin{cases}
\sqrt{4 \pi n_e}, & \nu < 1/2, \cr
\sqrt{4 \pi ({B \over \Phi_0} - n_e)}, & \nu > 1/2.
\end{cases}
\end{align}
The error in taking $k_F^\ast = \sqrt{4 \pi n_e}$ for $\nu>1/2$ is roughly 2\% at the $p=-1$ minimum; this error decreases for higher $|p|$ as $\nu=1/2$ is approached.
This result \eqref{kamburovresult}, which has been confirmed by subsequent measurements \cite{PhysRevLett.125.046601, PhysRevLett.117.096601, hossain2021bloch}, presents a challenge for the HLR theory.
There is no obvious reason for $k_F^\ast$ to vary about $\nu=1/2$.
One attempt \cite{BMF2015} for an explanation, along the lines suggested by the experiment, is to use the HLR theory for $\nu<1/2$ and to introduce an $a$ $priori$ different composite fermion theory of holes (with density $n_h = B/\Phi_0 - n_e$) for $\nu>1/2$, with a (rounded) transition between the two states at $\nu=1/2$.
Within a mean-field approximation, in which gauge field fluctuations are ignored, a detailed analysis \cite{2017PhRvX...7c1029W, PhysRevB.95.235424} shows that the two theories do not produce Weiss oscillations consistent with experiment.
What is more, the composite fermion theory of holes gives precisely the same oscillation minima as the HLR theory, both for $\nu>1/2$ {\it and} $\nu < 1/2$.

An alternative composite fermion theory for the $\nu=1/2$ metal, proposed by Son \cite{Son2015}, uses Dirac ``composite" fermions interacting via an emergent gauge field without Chern-Simons term (see also \cite{WangSenthilfirst2015, MetlitskiVishwanath2016, Seiberg:2016gmd, PhysRevX.6.031043}).
This Dirac composite fermion theory was also studied in Ref. \cite{PhysRevB.95.235424}, within the mean-field approximation, and found to agree to $.002\%$ accuracy (for the $n_e$ and $a$ relevant to experiment) with the predictions of HLR theory.
Going beyond mean-field theory, it was argued in Ref. \cite{PhysRevB.100.165122} that gauge field fluctuations catalyze a $B^\ast$-dependent Dirac mass away from $\nu=1/2$.
When this effect is included the comparison with experiment is dramatically improved (almost too well!), suggesting that asymmetrically-distributed minima $B^\ast_p$ may be an indirect signature of the emergent gauge field.

There are two criticisms of Ref. \cite{PhysRevB.100.165122}.
The first is that the analysis used a $1/N$ expansion, in which the Dirac composite fermion theory was generalized to a theory with $N$ identical flavors of fermions, for the (controlled) calculation of the fluctuation-induced mass.
While the $1/N$ expansion is standard (e.g., Ref. \cite{MOSHE200369}), there is no guarantee the large-$N$ result can be smoothly continued to or remain accurate at small $N$.

Second---and most relevant to this paper---the Dirac composite fermion theory used in Ref. \cite{PhysRevB.100.165122} lacks Galilean invariance.
(We remark that the composite fermion wave function approach does not suffer from this deficiency \cite{jainCF}.
The value of $k^\ast_F$, extracted from the oscillations in the pair-correlation functions of Jain states at $\nu=n/2n+1$ about $\nu=1/2$, appears to be consistent with the experimental result \eqref{kamburovresult} \cite{Balram2015,Balram2017}.)
This symmetry should be approximate in the electron system, if disorder is sufficiently weak.
The composite fermion theory considered in Ref. \cite{PhysRevB.100.165122} can be made to preserve Galilean invariance, only if additional correction terms are included  \cite{2013arXiv1306.0638T, sondiracreview, PhysRevB.97.195314}.
These correction terms are organized in an expansion in momentum $|{\bf q}|$, with coefficients fixed by symmetry.
Might these correction terms account for the discrepancy between Dirac composite fermion mean-field theory and experiment, without the need to invoke emergent gauge field fluctuations?
Here, we answer this question in the negative (see Fig.~\ref{rhooscillation}), using a composite fermion mean-field theory in which Galilean invariance is preserved to ${\cal O}(|{\bf q}|^2)$ (see Ref. \cite{PhysRevB.104.115401} for an analogous study of electromagnetic response).
To this order, the correction term is known as the dipole term because it involves a coupling of the composite fermion dipole moment to the external electromagnetic field.
The use of the Dirac composite fermion theory (rather than the HLR theory) allows for a direct comparison with the results in Ref. \cite{PhysRevB.95.235424, PhysRevB.100.165122}.
We find that, while an approximate Galilean invariance improves the comparison of theory with experiment, it is insufficient to fully explain the discrepancy (see Fig.~\ref{errorplot}).

We remark that the issue of the effects of an approximate Galilean symmetry on the Weiss oscillations produced by the Dirac composite fermion theory is independent of the issue of particle-hole symmetry about half-filling fraction \cite{girvin1984}.
The observed minima \cite{Kamburov2014, PhysRevLett.125.046601, PhysRevLett.117.096601, hossain2021bloch} and the predictions of various theoretical analyses \cite{2017PhRvX...7c1029W, PhysRevB.95.235424, PhysRevB.100.165122}, including that of this paper, are consistent with particle-hole symmetry.
To be consistent with particle-hole symmetry, the minima should appear symmetrically about half-filling fraction as a function of $\delta n_e = n_e - B/2\Phi_0$, with $B$ held fixed.
(Note that the experiments are performed at fixed $n_e$ and varying $B$; our statement regarding particle-hole symmetry assumes the locations of the oscillation minima continue to be described by Eqs.~\eqref{weissformula} and \eqref{kamburovresult} in a hypothetical experiment at fixed $B$ and varying $n_e$.) 
As we mentioned above, the $B^\ast$-dependent Dirac mass distinguishes Ref. \cite{PhysRevB.100.165122} from Ref. \cite{2017PhRvX...7c1029W, PhysRevB.95.235424}.
This mass has the form $m \sim {\rm sign}(B^\ast) |B^\ast|^{1/3} |B|^{1/6}$ and appears quadratically in the formula that determines the locations of the oscillation minima.
It seems that the asymmetric profile of this mass at fixed $n_e$ and varying $B$ is responsible for the improved agreement between theory and experiment.

Another motivation for this paper comes from a recent study of Weiss oscillations about $\nu = 1/4$ \cite{PhysRevB.100.041112}.
In contrast to $\nu=1/2$, the oscillation minima $B^\ast_p$ are symmetrically distributed about $\nu=1/4$ and consistent with $k_F^\ast = \sqrt{4 \pi n_e}$.
This difference in behavior, when compared to $\nu=1/2$, may be due to a lack of experimental resolution, since only $|p| \leq 2$ minima are resolvable, similar to the pre-2014 studies of the $\nu=1/2$ quantum oscillations.
To study if instead there might be a qualitatively different behavior of the $\nu=1/4$ metal, we study the Dirac composite fermion description of the $\nu=1/4$ state \cite{PhysRevB.99.125135, PhysRevB.98.165137, PhysRevLett.122.257203}, with and without approximate Galilean invariance.
In principle, an approximate Galilean symmetry should have greater relevance in the cleaner samples that feature the $\nu=1/4$ metal.
Nevertheless, in our mean-field analysis, we do not find qualitative differences in the predicted Weiss oscillations.
These analyses suggest that gauge field fluctuations should be included for a satisfactory description: mean-field theory is inadequate.

The remainder of this paper is organized as follows.
In Sec. \ref{diraccompositefermionmeanfieldtheorysection} we derive the Dirac composite fermion mean-field theory of the approximately Galilean-invariant $\nu=1/2m$ state.
In Sec. \ref{weissanalysis}, we use this theory to determine the effect of an approximate Galilean invariance on the Weiss oscillations of the $\nu=1/2m$ state.
In Sec. \ref{conclusion}, we conclude and discuss possible directions of future work.
In Appendix \ref{lagrangianappendix}, we show how the Dirac composite fermion theory for the $\nu=1/2m$ state with $m>1$, studied in Sec. \ref{diraccompositefermionmeanfieldtheorysection}, is obtained by applying modular transformations to the Galilean-invariant Dirac composite fermion theory of the $\nu=1/2$ state.

\section{Dirac Composite Fermion Mean-Field Theory}

\label{diraccompositefermionmeanfieldtheorysection}

In this section, we derive a Dirac composite fermion mean-field theory with approximate Galilean symmetry for the $\nu  = 1/2m$ state.
Our primary goal is to determine how an external scalar potential couples to the composite fermion degrees of freedom. 
Knowing this coupling will then enable us to calculate the predicted Weiss oscillations in the next section.

We begin with the following Dirac composite fermion Lagrangian for electrons at $\nu = 1/2m$: ${\cal L} = {\cal L}_0 + {\cal L}_{\rm dipole}$, with ($\hbar = c = e = 1)$
\begin{align}
\label{dirachalflagrangian}
{\cal L}_0 & = \bar \psi i \slashed{D} \psi - {m - 1 \over 2 m} {1 \over 4 \pi} a d a - {1 \over 2 m} {1 \over 2 \pi} a d A + {1 \over 2 m} {1 \over 4 \pi} A d A, \\
\label{dipoleterm}
{\cal L}_{\rm dipole} & = {\epsilon^{k j} {\cal E}_j \over 2 {\cal B}} \big[ \psi^\dagger i D_k \psi + \big(i D_k \psi \big)^\dagger \psi \big]. 
\end{align}
(See Appendix \ref{lagrangianappendix} for a derivation of this Lagrangian for $\nu=1/2m$, $m \neq 1$, from the corresponding $\nu=1/2$ Lagrangian.)
Here, ${\cal L}_0$ is the original Dirac composite fermion Lagrangian \cite{Son2015}, without Galilean invariance, and ${\cal L}_{\rm dipole}$ \cite{2013arXiv1306.0638T, sondiracreview} is the dipole term; the Dirac composite fermion is $\psi$ ($\bar \psi = \psi^\dagger \gamma^0)$; the covariant derivative $\slashed{D} = D_\alpha \gamma^\alpha = \big( \partial_\alpha - i a_\alpha \big) \gamma^\alpha$, where $\alpha \in \{t,x,y\} = \{0,1,2\}$ and the $\gamma$ matrices $(\gamma^0, \gamma^1, \gamma^2) = (\sigma^3, i \sigma^2, - i \sigma^1)$ (the standard Pauli matrices); 
$a_\alpha$ is a dynamical (2+1)-dimensional  Abelian gauge field and $A_\alpha$ is the external electromagnetic field;
Chern-Simons terms are defined as $A d B = \epsilon^{\alpha \beta \gamma} A_\alpha \partial_\beta B_\gamma$, where $\epsilon^{\alpha \beta \gamma}$ is the totally antisymmetric symbol with $\epsilon^{012} = 1$.
In this section, we rescale the Dirac composite fermion velocity $v_F = 1$; we will assume throughout that the external magnetic field $B$ is spatially uniform and time independent.
The fields ${\cal E}_j$ and ${\cal B}$ appearing in the dipole term are
\begin{align}
{\cal E}_j & = \partial_j c_0 - \partial_0 c_j, \\
{\cal B} & = \epsilon^{jk} \partial_j c_k, 
\end{align}
where $c_\alpha$ is the linear combination:
\begin{align}
c_\alpha & = {(m-1) \over m} a_\alpha + {1 \over m} A_\alpha.
\end{align}

The dipole term ensures that ${\cal L}$ has an approximate Galilean symmetry \cite{2013arXiv1306.0638T, sondiracreview}.
Under the transformation ${\bf x'} = {\bf x} - {\bf v}t$, $A_{0}'({\bf x'},t')=A_{0}({\bf x},t)+v^{i}A_{i}({\bf x},t)$, $A_{i}'({\bf x'},t')=A_{i}({\bf x},t)$, with an identical transformation for $a_\alpha$, the Lagrangian transforms into itself up to 
\begin{align}
\delta {\cal L} = \left( 1 - {B \over m \big( B - 4 \pi ( m - 1 ) \bar{n}_{e} \big) } \right) { v^{k} \over 2} \big[ \psi^\dagger i D_k \psi + \big(i D_k \psi \big)^\dagger \psi \big] + {\cal O}(v^{2}).
\end{align}
For $m =1$, $\delta {\cal L} = 0$ to ${\cal O}(v^2)$; for $m > 1$, there is a nonzero linear in $|v|$ term, whose coefficient is arbitrarily small for $\nu \rightarrow 1/2m$.
In this sense, we say that the Dirac composite fermion Lagrangian, \eqref{dirachalflagrangian} + \eqref{dipoleterm}, has approximate Galilean symmetry.

The dipole term introduces a nonminimal coupling, between the electromagnetic field $A_\alpha$ and the composite fermions, which produces a term involving the composite fermion dipole moment,
\begin{align}
d_j = {\epsilon^{k j} \over 2 {\cal B}} \big[ \psi^\dagger i D_k \psi + \big(i D_k \psi \big)^\dagger \psi \big],
\end{align}
in the expression for the electron density $n_e$:
\begin{align}
\label{electrondensity}
n_e = {\delta {\cal L} \over \delta A_0} = {1 \over 2 m} {1 \over 2 \pi} \big(B - b) - {1 \over m} \partial_j d_j.
\end{align}
We denote $b = \epsilon^{ij} \partial_i a_j$.

We now describe the mean-field approximation.
Since $a_0$ appears linearly in ${\cal L}$, we treat it as a Lagrange multiplier field, whose equation of motion sets
\begin{align}
\label{a0constraint}
\psi^\dagger \psi - {m-1 \over 2 m} {b \over 2 \pi}- {1 \over 2 m} {B \over 2\pi} - {m - 1 \over m} \partial_j d_j = 0.
\end{align}
Equations \eqref{electrondensity} and \eqref{a0constraint} together imply that the composite fermion density,
\begin{align}
\label{densityconstraint}
\psi^\dagger \psi = {B \over 4 \pi} - (m-1) n_e.
\end{align}
For $m=1$, this is the familiar constraint that the Dirac composite fermion density is fixed by the external magnetic field $B$.
For $m>1$, we see that the composite fermion density depends on both $B$ and the electron density.

In mean-field theory, $b = \epsilon^{ij} \partial_i a_j$ is taken to be uniform and time-independent.
(In principle $b$ could be an arbitrary function, specified by external parameters; the choice of constant $b$ is sufficient for our purposes.)
We claim that spatially uniform deviations away from $\nu = 1/2m$ are fully accounted for by nonzero $b$:
\begin{align}
\label{effectiveb}
b = B - 4 m \pi \bar n_e.
\end{align}
Here, $\bar n_e = {1 \over {\rm volume}} \int d^2 x\ \langle n_e \rangle$ is the spatially uniform part of the electron density \eqref{electrondensity}.
The dipole moment term ($\sim \partial_i d_i$) does not contribute to the uniform electron density, so long as the combination ${\cal P} {\cal T}$ of parity and time-reversal symmetries \cite{2014JHEP...12..138G, Son2015} \footnote{The ${\cal PT}$ transformation rules are: ${\cal P T} \psi(x) ({\cal PT})^{-1} = \sigma^3 \psi(x')$; ${\cal P T} a_0(x) ({\cal PT})^{-1} = a_0(-t, x, -y)$, ${\cal P T} a_x(t, x, y) ({\cal PT})^{-1} = - a_x(-t, x, -y)$, ${\cal P T} a_y(t, x, y) ({\cal PT})^{-1} = a_y(-t, x, -y)$; $A_\alpha$ transforms identically; ${\cal P T} i ({\cal PT})^{-1} = -i$.} is preserved:
\begin{align}
\label{uniformdipolecontribution}
{1 \over {\rm volume}} \int d^2 x \langle \partial_j d_j \rangle \sim {1 \over {\rm volume}} \int d^2 x \ \langle \epsilon^{ij} \partial_i \big[\psi^\dagger({\bf x}) D_j \psi({\bf x}) \big] \rangle = 0.
\end{align}
(${\cal B}$, which equals $B - 4 \pi (m-1) \bar n_e$ for uniform $b$, has been factored out of the the dipole contribution to the electron density.)
Under ${\cal PT}$ (at $t=0$), the expectation value transforms to $- \langle \epsilon^{ij} \partial_i \big(\psi^\dagger(x, - y) D_j \psi(x, - y) \big) \rangle$ and must vanish when integrated over all of space, even in the presence of nonzero $b$.
We can see explicitly that this integral vanishes when the composite fermions fill some number of Landau levels due to the presence of a uniform $b$ [in principle, different than $B - 4 m \pi \bar n_e$, if the integral \eqref{uniformdipolecontribution} is nonzero].
These Landau levels, filled by the composite fermion with respect to the effective magnetic field $b$, are sometimes known as Lambda levels \cite{jainCF}. 
Substituting the wave function $\psi_{n, k_y}$ of the $k_y$th state of a Dirac particle in the $n$th Landau level [obtained in the next section in \eqref{wavefunction}] for $\psi$, we find the integrand in \eqref{uniformdipolecontribution} to be equal to
\begin{align}
\epsilon^{ij} \partial_i \big(\psi^\ast_{n} D_j \psi_{n} \big) \sim i \partial_x \left( \big( k_y - {x \over \ell_b^2} \big) H_n^{2}\big( {x  \over \ell_b} + k_y \ell_b \big) e^{- (x/\ell_b+k \ell_b)^2 } \right).
\end{align}
Integrating this produces a boundary term that vanishes as $|x| \rightarrow \infty$ for any fixed $k$.

While dipole term does not contribute to the uniform electron density (for unbroken ${\cal PT}$), it is responsible \cite{PhysRevB.97.195314} for an approximate Girvin-MacDonald-Platzman (GMP) \cite{PhysRevB.33.2481} algebra obeyed by the Fourier transformed electron densities $n_e({\bf q})$.
When $m=1$ (i.e., for $\nu = 1/2$), the densities $n_e({\bf q})$ satisfy the linearized GMP algebra:
\begin{align}
\label{linearizedGMPalgebra}
\big[n_e({\bf q}), n_e({\bf q}') \big] = i \ell_B^2 \epsilon^{jk} q_j q_k n_e({\bf q} + {\bf q}'),
\end{align}
appropriate for a Galilean-invariant system of electrons projected to the lowest Landau level.
This algebra follows from the canonical anticommutation relations for $\psi$ (and $\psi^\dagger$) and uses \eqref{densityconstraint}.
That the GMP algebra is only satisfied to ${\cal O}(|{\bf q}|^2)$ implies that additional terms must supplement the Lagrangian, if it is to have an exact Galilean symmetry.
When $m>1$ (e.g., $\nu = 1/4$), the Fourier-transformed densities $n_e({\bf q})$ do not obey \eqref{linearizedGMPalgebra}.
Instead, \eqref{linearizedGMPalgebra} is modified by replacing $\ell_B^2 \rightarrow \ell_{m {\cal B}}^2$ with ${\cal B} = B - 4 \pi (m-1) \bar n_e$.
Nevertheless, for $(m-1) | 1 - 2 m \nu | \ll 1$, the linearized algebra is satisfied to good approximation.
Therefore, at the cost of some abuse of description, we will consider the Dirac composite fermion theory ${\cal L}$ to have an approximate Galilean symmetry [in the sense of \eqref{linearizedGMPalgebra}] for any $m \geq 1$.

We now present the mean-field Lagrangian.
Into the starting Lagrangian \eqref{dirachalflagrangian} + \eqref{dipoleterm}, we introduce a chemical potential $\mu$ for $\psi^\dagger \psi$ to satisfy \eqref{densityconstraint} on average, set $a_0 = 0$, and finally substitute $a_j \rightarrow \bar a_j$, where $\epsilon^{jk} \partial_j \bar a_k = b$, with the uniform value of $b$ given in \eqref{effectiveb}.
The result is
\begin{align}
\label{generalpmeanfield}
{\cal L}_{\rm mf} & = \bar \psi i \bar D_\alpha \gamma^\alpha \psi + {1 \over 2 m} {\epsilon^{k j} \partial_j A_0 \over \big( B - 4 \pi (m - 1) \bar n_e \big)} \big[ \psi^\dagger i \bar D_k \psi + (i \bar D_k \psi)^\dagger \psi \big],
\end{align}
where $(\bar D_0, \bar D_j) \equiv (\partial_0 - i \mu, \partial_j - i \bar a_j)$.
In this mean-field Lagrangian, we have dropped terms that only involve external fields.
It is interesting to contrast the coupling between $A_0$ and the composite fermion dipole moment with what occurs in Dirac composite fermion theory, without the dipole term.
In the latter, the external scalar potential sources an additional contribution to the vector potential $\delta \bar a_j \sim \epsilon^{jk} \partial_k A_0$, which, in turn, couples to the Dirac composite fermion current.
(This follows from the $a_j$ equation of motion $\delta {\cal L}_0/\delta a_j = 0$, without dipole term.)
In the next section, we will show that, despite these apparently different couplings to $A_0$, the predicted Weiss oscillations are surprisingly similar.

\section{Weiss Oscillations}

\label{weissanalysis}

In this section, we study the quantum oscillations in the dc electrical resistivity $\Delta \rho_{xx}$ about $\nu=1/2m$ due to the periodic scalar potential,
 \begin{align}
 A_0 = V_0 \cos(K x), \quad K = 2\pi/a.
\end{align}
See, e.g., Ref. \cite{Kamburov2014}, for details on how this potential is implemented experimentally.
In Dirac composite fermion mean-field theory,
\begin{align}
\label{conductivitydictionary}
\Delta \rho_{xx} \propto \Delta \sigma_{yy}^{\psi},
\end{align}
where $\Delta \sigma_{yy}^{\psi}$ is the correction to the Dirac composite fermion conductivity due to $A_0$ (see Ref. \cite{PhysRevB.95.235424} for a derivation of this relation).
There are generally oscillatory corrections to $\rho_{yy}$ and $\rho_{xy}$, however, their amplitudes are typically less prominent and so we concentrate on $\Delta \rho_{xx}$.
Finite frequency and wave vector corrections to the formula \eqref{conductivitydictionary}---when the dipole term is present---have been computed by Hofmann \cite{PhysRevB.104.115401} in the random phase approximation.

We will use the Kubo formula \cite{charbonneauvVV1982} to find $\Delta \sigma_{yy}^{\psi}$:
\begin{align}
\label{kubo}
\Delta \sigma_{ij}^\psi = {1 \over L_x L_y} \Sigma_M \big( \partial_{E_M} f_D(E_M) \big) \tau(E_M) v_i^M v_j^M,
\end{align}
where $L_x$ ($L_y$) is the linear system size along the $x$ direction ($y$ direction), the sum $\Sigma_M$ is over single-particle states with quantum number $M$, $\tau(E_M)$ is the scattering time for states with energy $E_M$, $f^{-1}_D(E) = 1 + \exp[\beta(E - \mu)]$ is the Fermi-Dirac distribution function for chemical potential $\mu$ and temperature $\beta^{-1} = T$, and $v_i^M$ is the velocity correction in the $x_i$-direction of the state $M$. 
We assume constant $\tau(E) = \tau > 0$. 
(We interpret this as a ``weak" breaking of spatial translation invariance.)
To calculate \eqref{kubo}, we will determine how the single-particle composite fermion energies $E_M$ are affected by $A_0$.
This, in turn, will determine the velocities $v_i^M$.
The leading correction in $V_0$ will be to $v_y^M$.
This is the reason we wrote $\Delta \sigma_{yy}^{\psi}$ in \eqref{conductivitydictionary}.

The Galilean-invariant Dirac composite fermion mean-field Hamiltonian, following from \eqref{generalpmeanfield}, is $H=H_{0}+H'$, with
\begin{eqnarray}
H_{0}&=&v_{F}\sigma^{j}(i\partial_{j}+\bar{a}_{j}),\\
H'&=&{1 \over 2 m} {i \epsilon^{ij} \partial_i A_0 \over \big(B - 4 \pi (m-1) \bar n_e \big)} \big( \psi^\dagger \bar D_j \psi - (\bar D_j \psi^\dagger) \psi \big),
\end{eqnarray}
where $\bar{a}_{j}=(0, bx)$ and we have reinstated the Fermi velocity $v_F$.
The effective magnetic field $b = B - 4 \pi m \pi \bar n_e$.
$H_{0}$ describes a relativistic particle placed in a uniform magnetic field $b$, with single-particle eigenstates:
\begin{align}\label{wavefunction}
\psi_{n,k_{y}}=\frac{e^{ik_{y}y}}{\sqrt{2L_{y}\ell_{b}}}
\begin{pmatrix}
-i\Phi_{n-1}\left(\frac{x+x_{b}}{\ell_{b}}\right)\\
\Phi_{n}\left(\frac{x+x_{b}}{\ell_{b}}\right)
\end{pmatrix},
\end{align}
where
\begin{align}\label{wavefunction-2}
\Phi_{n}(x)=\frac{e^{-x^{2}/2}}{\sqrt{2^{n}n!\sqrt{\pi}}}H_{n}(x).
\end{align}
Here, $n$ is a nonnegative integer, $H_{n}(x)$ is the $n$-th order Hermite polynomial, and $k_y$ is the wave vector along the $y$-direction. 
(We define $\Phi_{-1} = 0$.)
The energies of these states are $E_{n,k_{y}}^{(0)}=\pm\sqrt{2n}v_{F}/\ell_{b}$.

First-order perturbation theory applied to $H'$ gives the energy level corrections:
\begin{eqnarray}
E_{n,k_{y}}^{(1)}&=&\frac{2}{2m}\frac{V_{0} b}{[B-4\pi (m-1)\bar{n}_{e}]}e^{-z/2}\cos(Kx_{b})\left[nL_{n}(z)-L_{n-1}(z)-(n-1)L_{n-2}(z)\right]\\
&&-\frac{1}{2 m}\frac{V_{0}K b}{[B-4\pi (m-1)\bar{n}_{e}]}e^{-z/2}\left[\frac{z}{K}\cos(Kx_{b})+x_{b}\sin(Kx_{b})\right]\left[L_{n}(z)+L_{n-1}(z)\right], \nonumber
\end{eqnarray}
where $x_{b}=k_{y}\ell_{b}^{2}$, and $L_{n}(z)$ with $z=K^{2}\ell_{b}^{2}/2$ is the $n$-th order Laguerre polynomial. 
Near $\nu = 1/2m$, $n \rightarrow \infty$ and we have
\begin{align}
E_{n,k_{y}}^{(1)}\approx\frac{4}{2 m}\frac{V_{0} b n}{[B-4\pi (m-1)\bar{n}_{e}]}\cos(Kx_{b})e^{-z/2}\left[L_{n}(z)-L_{n-1}(z)\right].
\end{align}
The leading contribution to the velocity is obtained via the semi-classical approximation,
\begin{align}
\Delta v_{y}^{n,k_{y}}=\frac{\partial E_{n,k_{y}}^{(1)}}{\partial k_{y}}=-\frac{4}{2 m}\frac{V_{0} bK\ell_{b}^{2}}{[B-4\pi (m-1)\bar{n}_{e}]}e^{-z/2}\sin(Kx_{b})n\left[L_{n}(z)-L_{n-1}(z)\right].
\end{align}
Notice there is no correction to $v_x$.

Hence, we have the correction to the conductivity induced by $H'$: 
\begin{align}
\Delta\sigma^{\psi}_{yy}=\frac{16}{(2 m)^{2}}\frac{\widetilde{\tau}K^{2}\ell_{b}^{2}V_{0}^{2} b^{2}}{[B-4\pi (m-1) \bar{n}_{e}]^{2}}\sum_{n=0}^{\infty}\frac{\beta g(E_{n,k_{y}}^{(0)})}{\left[1+g(E_{n,k_{y}}^{(0)})\right]^{2}}\left[ne^{-z/2}L_{n}(z)-ne^{-z/2}L_{n-1}(z)\right]^{2},
\end{align}
with $g(E)=e^{\beta(E-E_{F})}$, where the Fermi energy $E_{F}=v_{F}k_{F}=v_{F}\sqrt{B-4 \pi (m-1) \bar{n}_{e}}$. 
For $n\rightarrow \infty$, the Laguerre polynomial can be expanded as
\begin{align}
e^{-z/2}L_{n}(z)\rightarrow \frac{\cos(2\sqrt{nz}-\frac{\pi}{4})}{\left(\pi^{2}nz\right)^{1/4}}+\frac{1}{16}\frac{\sin(2\sqrt{nz}-\frac{\pi}{4})}{[\pi^{2}(nz)^{3}]^{1/4}}+\mathcal{O}(\frac{1}{n^{5/4}}).
\end{align}
We note that this expansion is of one higher-order in the expansion in $1/n$ than required when $A_0$ induces a scalar or vector potential perturbation to the Dirac composite fermion mean-field Hamiltonian, as in, e.g., Refs. \cite{PhysRevB.95.235424, PhysRevB.100.165122}.
The continuum limit is taken by substituting
\begin{align}
n\rightarrow\frac{E^{2}\ell^{2}_{b}}{2v_{F}^{2}},\ \sum_{n}\rightarrow\frac{\ell^{2}_{b}}{v_{F}^{2}}\int\limits_{-\infty}^{+\infty}EdE.
\end{align}
Thus, $\Delta\sigma^{\psi}_{yy}$ becomes
\begin{eqnarray}
\Delta\sigma^{\psi}_{yy}&=&\mathcal{A}\int\limits_{-\infty}^{+\infty}dE\frac{\beta g(E)}{\left[1+g(E)\right]^{2}}\sin^{2}\left( \frac{K\ell_{b}^{2}E}{v_{F}}-\delta-\frac{\pi}{4}\right),\\
\mathcal{A}&=&\frac{2\tilde{\tau}K\ell_{b}^{4}V_{0}^{2}E_{F}^{3}}{\pi(2m)^{2}(B-4\pi (m-1) \bar{n}_{e})^{2}v_{F}^{3}}\sec^{2}\left(\delta\right)\sin^{2}\left(\frac{Kv_{F}}{E_{F}}\right),
\end{eqnarray}
where $\delta=\tan^{-1}(v_{F}/8K\ell_{b}^{2}E_{F})$. 
For low temperatures $\beta^{-1}\ll E_{F}$, we write $E\approx E_{F}+s\beta^{-1}$ so that $\Delta\sigma^{\psi}_{yy}$ becomes
\begin{eqnarray}
\label{oscillation}
\Delta\sigma^{\psi}_{yy}&=&\mathcal{A}\int\limits_{-\infty}^{+\infty}ds\frac{e^{s}}{\left(1+e^{s}\right)^{2}}\sin^{2}\left( \frac{K\ell_{b}^{2}}{\beta v_{F}}s+\frac{K\ell_{b}^{2}E_{F}}{v_{F}}-\delta-\frac{\pi}{4}\right)\nonumber\\
&=&\mathcal{A}\left\{ \frac{1}{2}\left[1-\frac{T/T_{0}}{\sinh(T/T_{0})}\right]+\frac{T/T_{0}}{\sinh(T/T_{0})}\sin^{2}\left(\frac{K\ell_{b}^{2}E_{F}}{v_{F}}-\delta-\frac{\pi}{4}\right)\right\},
\end{eqnarray}
where 
\begin{align}
T_{0}^{-1}=\frac{2\pi K\ell_{b}^{2}}{v_{F}}.
\end{align}
\begin{figure}[t] 
\includegraphics[height=4in,width=7in] {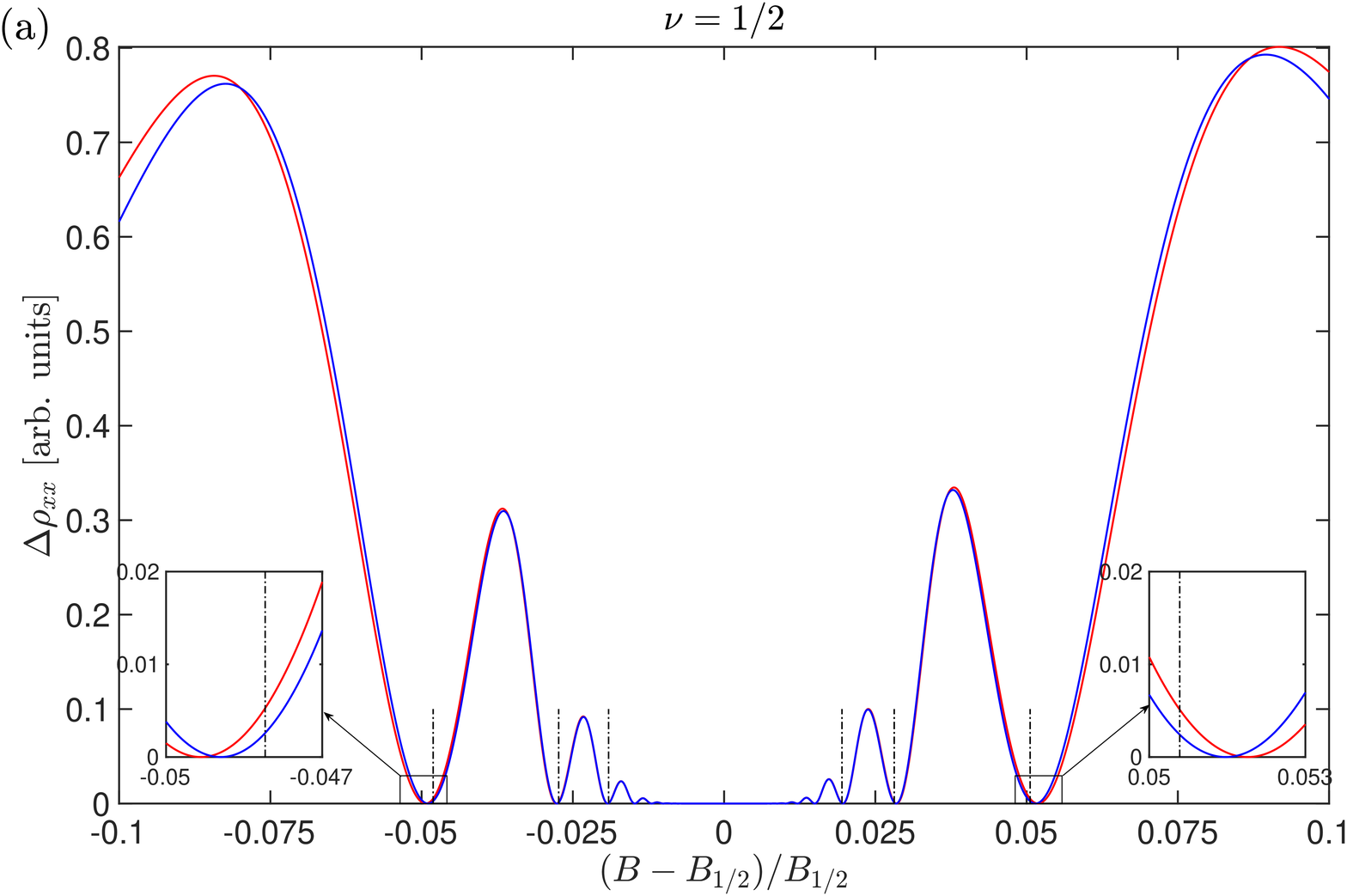}
\includegraphics[height=4in,width=7in] {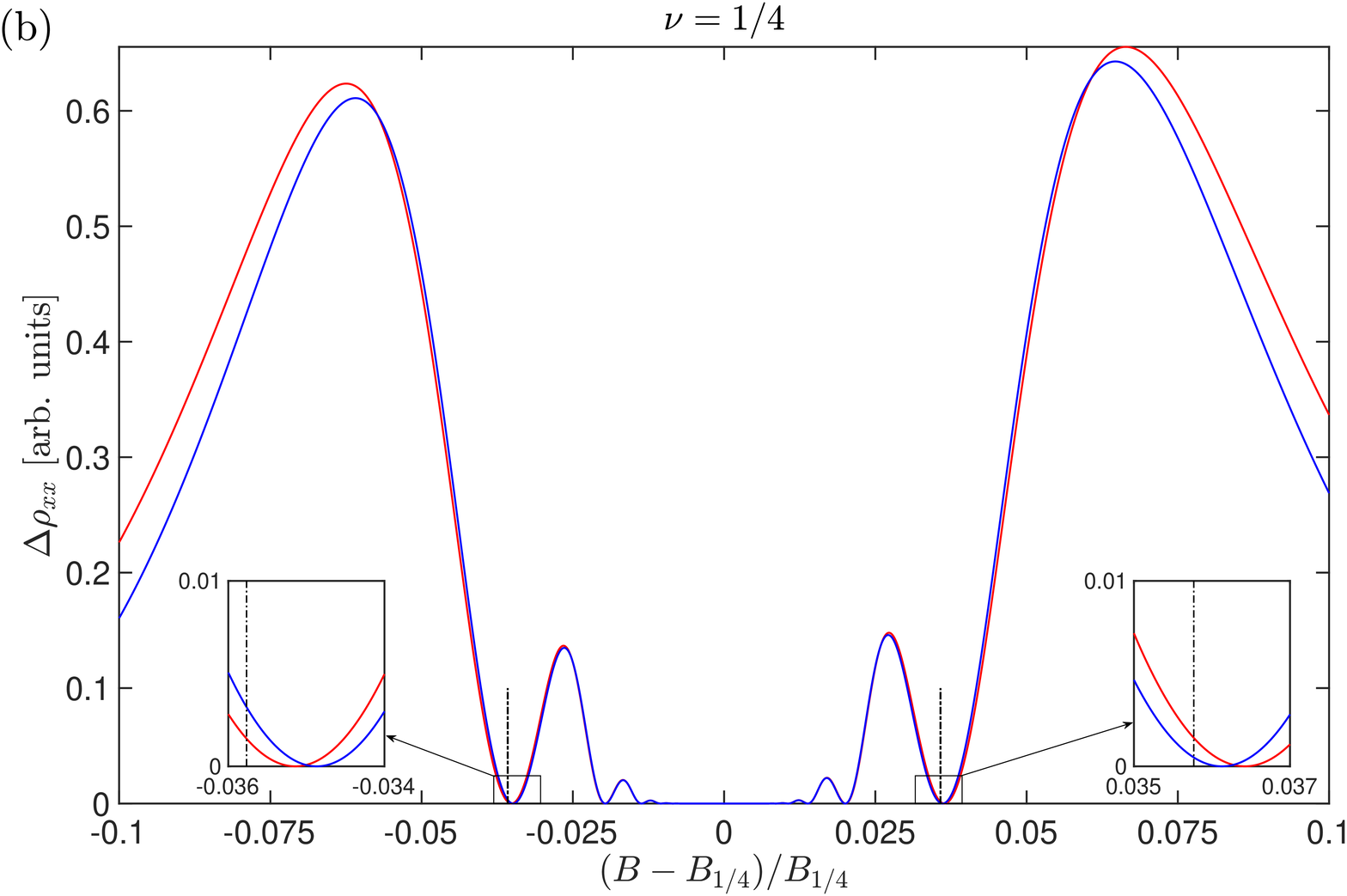}
\centering
\caption{(a) The oscillatory part of $\rho_{xx}$ near half-filling: $B_{a}=\hbar c/ea^{2}$, $T=0.06\sqrt{2B_{1/2}}$, and $B_{a}/B_{1/2}=10^{-3}$. (b) The oscillatory part of $\rho_{xx}$ near quarter filling: $B_{a}=\hbar c/ea^{2}$, $T=0.06\sqrt{2B_{1/4}}$, and $B_{a}/B_{1/4}=10^{-3}$. The blue(red) curves are the predictions of Dirac composite fermion mean-field theory with (without) dipole term; the hash marks approximate the oscillation minima locations found experimentally.} 
\label{rhooscillation}
\end{figure}
We plot the oscillatory part of $\Delta \sigma^\psi_{yy} \sim \Delta \rho_{xx}$ in Fig.~\ref{rhooscillation}.

The oscillation minima in \eqref{oscillation} occur at
\begin{align}
\frac{K\ell_{b}^{2}E_{F}}{v_{F}}-\tan^{-1}\left(\frac{v_{F}}{8K\ell_{b}^{2}E_{F}}\right)-\frac{\pi}{4}=p\pi.
\end{align}
Solving this condition in terms of the effective magnetic field $b$, we find the $p$th oscillation minima $b_p$:
\begin{align}
b_{p}=
\begin{cases}
\left( \frac{ \sqrt{ B_{a} } }{ ( p + \frac{1}{4} ) + \frac{1}{ 8 \pi^2 ( p + \frac{1}{4} ) } } + \sqrt{ \left ( \frac{ \sqrt{ B_{a} } }{ ( p + \frac{1}{4} ) + \frac{1}{ 8 \pi^2 ( p + \frac{1}{4} ) } } \right)^{2} + { 1 \over m } B_{ \frac{1}{ 2 m } } } \right)^{2} - {1 \over m} B_{ \frac{1}{2m} }, & \nu < {1 \over 2m}\cr
\left( - \frac{ \sqrt{ B_{a} } }{ ( p + \frac{1}{4} ) + \frac{1}{ 8 \pi^2 ( p + \frac{1}{4} ) } } + \sqrt{ \left( \frac{ \sqrt{B_{a} } }{ ( p + \frac{1}{4} ) + \frac{1}{8 \pi^2 ( p + \frac{1}{4} ) } } \right)^{2} + {1 \over m} B_{ \frac{1}{2m} } } \right)^{2} - {1 \over m} B_{ \frac{1}{2m} }, & \nu > {1 \over 2m}
\end{cases}
,
\label{oscillationresult}
\end{align}
where $B_{a}=\hbar c/e a^{2}$ is the magnetic field defined by the period $a$ of the potential $A_0$ and $B_{{1 \over 2m}}$ is the magnetic field at $\nu=1/2m$. 
For reference, we compare the result \eqref{oscillationresult} to the predicted oscillation minima $b_p'$ of Dirac composite-fermion mean-field theory without dipole term \cite{PhysRevB.95.235424}:
\begin{align}
b'_{p}=
\begin{cases}
\left( \frac{ \sqrt{ B_{a} } }{ ( p + \frac{1}{4} ) } + \sqrt{ \left( \frac{ \sqrt{ B_{a} } }{ ( p + \frac{1}{4} ) } \right)^{2} + {1 \over m} B_{\frac{1}{2m}}}\right)^{2} - {1 \over m} B_{ \frac{1}{2m} }, & \nu<{1 \over 2m}\cr
\left( - \frac{ \sqrt{ B_{a} } }{ ( p + \frac{1}{4} ) }+\sqrt{ \left( \frac{ \sqrt{ B_{a} } }{ ( p + \frac{1}{4} ) } \right)^{2} + {1 \over m} B_{ \frac{1}{2m} } }\right)^{2} - {1 \over m} B_{ \frac{1}{2m} }, & \nu>{1 \over 2m}
\end{cases}
.
\label{diracwithout}
\end{align}
We see that the effect of the dipole term weakens for increasing $p$.

\begin{figure}[t] 
\includegraphics[height=3in,width=4in] {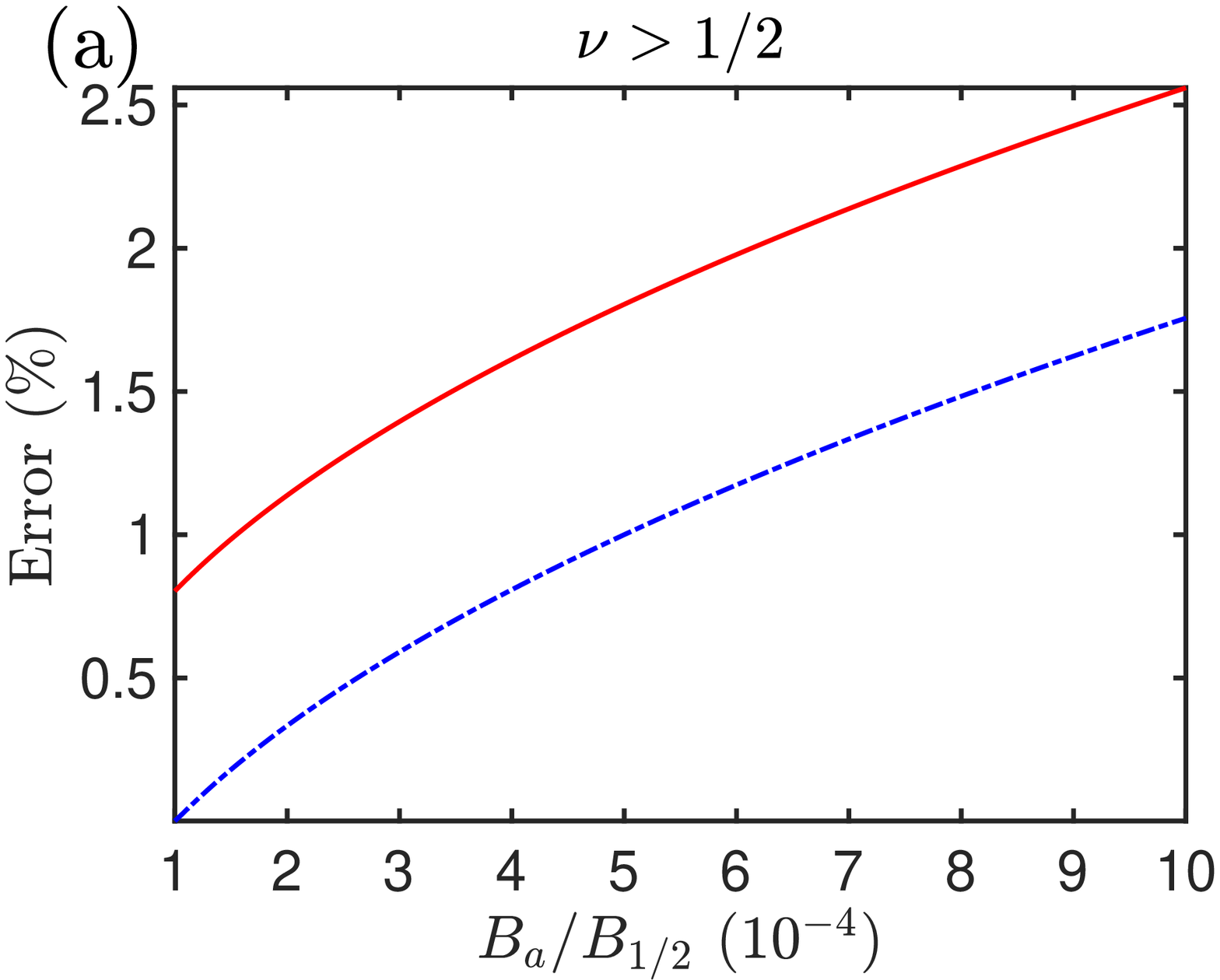}
\includegraphics[height=3in,width=4in] {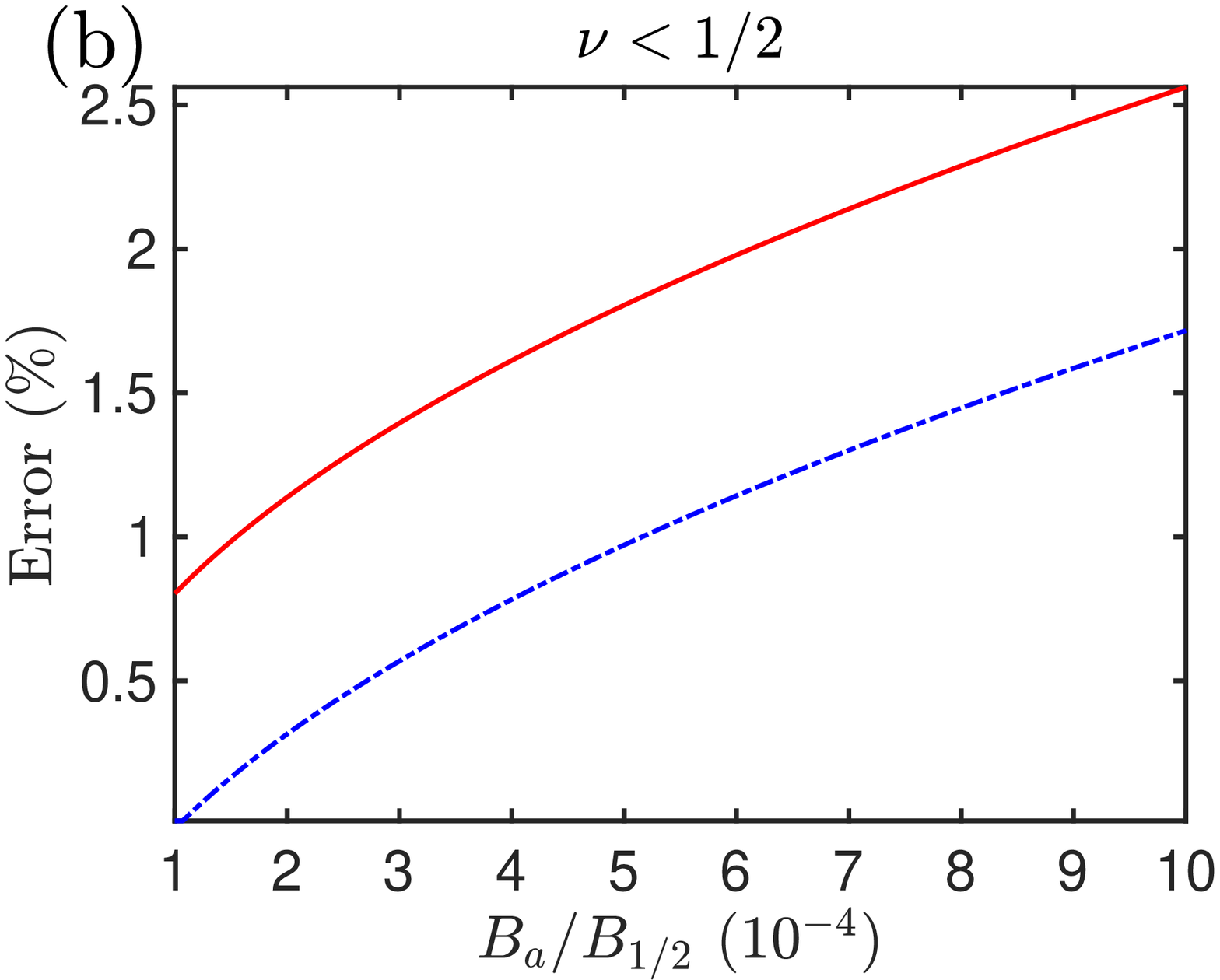}
\centering
\caption{Error between the predicted $|p|=1$ oscillation minima of Dirac composite fermion-mean-field theory with (blue curve) and without (red curve) the dipole term and the experimental minima \eqref{experimentalresults} as a function of $B_a/B_{1/2}$  for (a) $\nu > 1/2$ and (b) $\nu < 1/2$.
The error is defined as $|b_{p} - B^\ast_{p}|/B^\ast_{p}$ for the minima $b_p$ in \eqref{oscillationresult}; the error for minima in \eqref{diracwithout} is the same with the replacement $b_p \rightarrow b_p'$.}
\label{errorplot}
\end{figure}

We may also compare \eqref{oscillationresult} with the approximate locations of the oscillation minima found experimentally.
For this, we set $\nu = 1/2$.
Solving \eqref{weissformula} for $B_p^{\ast}$ with $k_F^\ast$ given by \eqref{kamburovresult}, we find
\begin{align}
B_{p}^{\ast}=
\begin{cases}
\frac{ 2 \sqrt{ B_{a} B_{1/2} } }{ ( p + \frac{1}{4} ) }, & \nu < {1 \over 2}\cr
{1 \over 2} \left( - \frac{ 2 \sqrt{ B_{a} } }{ ( p + \frac{1}{4} ) } + \sqrt{ \left( \frac{ 2 \sqrt{ B_{a} } }{ ( p + {1 \over 4} ) }\right)^{2} + B_{ 1 \over 2 } }\right)^{2} - {1 \over 2} B_{1 \over 2}, & \nu > {1 \over 2}
\end{cases}
.
\label{experimentalresults}
\end{align}
In Fig.~\ref{errorplot}, we plot the error, i.e., the percentage difference between the mean-field predictions \eqref{oscillationresult} and \eqref{diracwithout} about $\nu=1/2$ and the experimental results \eqref{experimentalresults} for the $|p|=1$ minima as a function of $B_a/B_{1/2}$.
We observe the small improvement provided by the dipole term. 
Notice that the error can be increased (decreased) by decreasing (increasing) the oscillation period $a$ at fixed $B_{1/2}$.

\section{Conclusion}

\label{conclusion}

In this paper, we studied the effect of an approximate Galilean symmetry on the quantum oscillations about even denominator filling fractions $\nu=1/2m$ of the lowest Landau level due to an applied periodic scalar potential.
For our analysis, we used the Dirac composite fermion theory.
Galilean symmetry requires that additional terms be included in the standard composite fermion Lagrangian; we considered the leading term, known as the dipole term.
In a mean-field approximation, we found that the dipole term causes a shift of the locations of the oscillation minima relative to their locations predicted by the theory without dipole term (see Fig.~\ref{rhooscillation}).
This shift improves the comparison of theory with experiment, however, it does not resolve the disagreement (see Fig.~\ref{errorplot}).

While we used the Dirac composite fermion theory to study the effect of an approximate Galilean symmetry, a complementary question is to determine the effect of the corresponding corrections in the HLR theory \cite{PhysRevB.54.R11114, Stern1999}.
Without these Galilean corrections, the HLR and Dirac theories agree to high degree, for the electron densities and potential periods considered experimentally. 
Whether or not this continues when Galilean corrections are included is an open question.

The Dirac composite fermion theory, with dipole term, that we used to study the $\nu=1/2$ state preserves Galilean symmetry to ${\cal O}(v^2)$, where $v$ is the velocity parameter of a Galilean boost \cite{2013arXiv1306.0638T, sondiracreview}.
This is discussed in Sec. \ref{diraccompositefermionmeanfieldtheorysection}.
The Dirac composite fermion theory for the $\nu=1/2m$ state, with $m >1$, that we derived in Appendix \ref{lagrangianappendix} preserves Galilean symmetry to ${\cal O}(c |v|)$, where the constant $c \sim (1 - 2 m \nu)$.
It would be interesting to engineer a theory for the $\nu=1/2m$ state that is invariant to ${\cal O}(v^2)$.

Mean-field theory ignores the fluctuations of the emergent gauge field, present in Dirac (and other) composite fermion theories.
Some fluctuation effects associated with this gauge field on quantum oscillations due to an applied scalar potential were considered in Ref. \cite{PhysRevB.100.165122}.
There it was found that the exchange of gauge bosons with $|{\bf q}| \leq |q_0|$ produced a magnetic field dependent Dirac mass away from $\nu = 1/2$.
Taking this mass term to be the leading fluctuation-correction to the Dirac composite fermion mean-field Hamiltonian, the resulting corrections to the oscillation minima were then computed and found to agree well with experiments.
Our results in this paper support the conclusion that gauge field effects are important for resolving the discrepancy between composite fermion theory and experiment.
It would be interesting to improve the calculation in Ref. \cite{PhysRevB.100.165122} by considering the effects of the exchange of gauge fields with $|q_0| <  |{\bf q}|$.
In this regime, Landau damping of the ``magnetic" component of the gauge field propagator is expected to result in IR dominant Dirac composite fermion self-energy corrections \cite{SSLee2009OrderofLimits, MetlitskiSachdev2010Part1, Mross2010}.
In particular, it would be interesting to understand this regime when/if a dynamically-generated Chern-Simons term for the gauge field is present.
These studies are expected to be sensitive to the nature of the electron-electron interactions. 

Without the dipole term, Dirac composite fermion mean-field theory can be argued to obtain in the limit of an infinite strength Coulomb interaction.
This is because the electron density is parameterized by the flux of the emergent gauge field in the theory without dipole term and so, in the limit of an infinite strength Coulomb interaction, fluctuations in the gauge field are suppressed.
When the dipole term is included, the electron density \eqref{electrondensity} is modified and, if \eqref{densityconstraint} is imposed, the dependence on the gauge field disappears.
The physical meaning of the mean-field approximation becomes less clear.

\section*{Acknowledgments}

We thank Hart Goldman, Shafayat Hossain, Pak Kau Lim, Sri Raghu, and Mansour Shayegan for useful conversations.
This material is based upon work supported by the U.S. Department of Energy, Office of Science, Office of Basic Energy Sciences under Award No.~DE-SC0020007.
P.K. is supported by NSF through the Princeton University (PCCM) Materials Research Science and Engineering Center DMR-2011750.

\appendix

\section{Derivation of the $\nu=1/2m$ Lagrangian}
\label{lagrangianappendix}

In this appendix, we derive the $\nu=1/2m$, with $m > 1$, Dirac composite fermion theory \eqref{dirachalflagrangian} + \eqref{dipoleterm} from the $\nu =1/2$ theory, ${\cal L}_1 + {\cal L}_2$, with
\begin{align}
\label{seed}
{\cal L}_1 & = i \bar{\psi} \not \! \! D \psi - {1 \over 2} {1 \over 4 \pi} a d a 
+ {1 \over 2 \pi} a d \tilde b - {2 \over 4 \pi} \tilde b d \tilde b - {1 \over 2 \pi} \tilde b d A, \\
\label{seeddipole}
{\cal L}_2(A, a) & = {\epsilon^{k j} E_j \over 2 B} \big( \psi^\dagger i D_k \psi + \big(i D_k \psi \big)^\dagger \psi \big),
\end{align}
where $E_j = \partial_j A_0 - \partial_0 A_j$ and $B = \epsilon^{ij} \partial_i A_j$.
Our conventions are the same as those given below \eqref{dirachalflagrangian} + \eqref{dipoleterm}.
Note that $\tilde b_\alpha$ is an Abelian gauge field and we now indicate the dependence of $A_\alpha$ and $a_\alpha$ in the dipole term.
Because $\tilde b$ appears quadratically, we may integrate it out by solving its equation of motion. 
The result is \eqref{dirachalflagrangian} + \eqref{dipoleterm} at $\nu = 1/2$.

The usefulness of the form of the Lagrangian \eqref{seed} + \eqref{seeddipole} is that Chern-Simons terms are properly normalized (assuming conventional quantization conditions on the gauge fields $a_\alpha$ and $b_\alpha$) and we may therefore apply modular transformations \cite{WittenSL2Z, Seiberg:2016gmd} to generate new Dirac composite fermion Lagrangians for other metallic states.
These modular transformations are defined as follows:
\begin{align}
\label{modulardef}
& {\bf T}: {\cal L}(\Phi, A) \mapsto {\cal L}(\Phi, A) + {1 \over 4 \pi} A d A, \\
& {\bf S}: {\cal L}(\Phi, A) \mapsto {\cal L}(\Phi, c) - {1 \over 2 \pi} c d B.
\end{align}
Above, ${\cal L}(\Phi, A)$ represents a general Lagrangian with dynamical field $\Phi$ and $U(1)$ symmetry, whose current couples to the external $U(1)$ field $A_\alpha$; the ${\bf S}$ transformation means that we first make the external field $A_\alpha$ dynamical by replacing it with the new Abelian gauge field $c_\alpha$ and we then introduce a new external gauge field $B_\alpha$ that couples to the (conserved) flux of $c_\alpha$ through a ``BF term."
Following Ref. \cite{PhysRevB.99.125135}, we obtain a Dirac composite fermion theory for the $\nu=1/2m$ state by applying ${\bf S}^{-1} {\bf T}^{- 2m + 2} {\bf S}$ to \eqref{seed} + \eqref{seeddipole}.
See also the related works \cite{PhysRevB.98.165137, PhysRevLett.122.257203}.
We find ${\cal L}_1 \rightarrow {\cal L}_1' = {\bf S}^{-1} {\bf T}^{- 2m + 2} {\bf S} {\cal L}_1$ and ${\cal L}_2 \rightarrow {\cal L}_2' = {\bf S}^{-1} {\bf T}^{- 2m + 2} {\bf S} {\cal L}_2$, with
\begin{align}
\label{nuoneover2m}
{\cal L}'_1 & = \bar \psi i \slashed{D} \psi - {1 \over 2} {1 \over 4 \pi} a d a + {1 \over 2 \pi} a d \tilde b - {2 \over 4 \pi} \tilde b d \tilde b - {1 \over 2 \pi} \tilde b d c \cr
& - {1 \over 2 \pi} c d g - {2 m - 2 \over 4 \pi} g d g + {1 \over 2 \pi} g d A, \\
{\cal L}'_2(c, a) & = - {\epsilon^{i j} {\cal E}_i \over 2 {\cal B}} \big( \psi^\dagger i D_j \psi - i \big(D_j \psi \big)^\dagger \psi \big), 
\end{align}
where ${\cal E}_i = \partial_i c_0 - \partial_0 c_i$, and ${\cal B} = \epsilon_{ij} \partial_i c_j$.
Application of the two ${\bf S}$ transformations has produced two new $U(1)$ gauge fields $c_\alpha$ and $g_\alpha$.

The Lagrangian \eqref{dirachalflagrangian} + \eqref{dipoleterm} is a simplified version of ${\cal L}_1' + {\cal L}_2'$ that obtains by approximately solving the equations of motion for the gauge fields $b_\alpha,c_\alpha,g_\alpha$.
Note here and below we are leaving the common vector subscript $\alpha$ for these fields implicit whenever convenient.
The equations of motion for $b, c, g$ are (up to gauge equivalence)
\begin{align}
a - c - 2 b & = 0, \\
- b - g + \partial_c {\cal L}_2' & = 0, \\
- c - (2m - 2) g + A & = 0.
\end{align}
The first and third equations are solved by taking (for $m \neq 1$)
\begin{align}
b = {a - c \over 2}, \\
\label{gsolution}
g = {A - c \over 2 m - 2}.
\end{align}
We approximately solve the $c$ equation by dropping the dipole term $\partial_c {\cal L}_2'$:
\begin{align}
\label{csolution}
c = {(2 m - 2) a + 2 A \over 2 m}.
\end{align}
The argument for this approximation is that the dropped term is of higher order than the term retained, in a $1/B$ expansion in the external magnetic field $B$.
Substituting these solutions into ${\cal L}_1' + {\cal L}_2'$ , we find \eqref{dirachalflagrangian} + \eqref{dipoleterm}.

\bibliography{bigbib}

\end{document}